\documentclass[aps,twocolumn]{revtex4}

\usepackage{graphicx}

\begin{document}
\newtheorem{theorem}{Theorem}
\newtheorem{acknowledgement}[theorem]{Acknowledgement}
\newtheorem{algorithm}[theorem]{Algorithm}
\newtheorem{axiom}[theorem]{Axiom}
\newtheorem{claim}[theorem]{Claim}
\newtheorem{conclusion}[theorem]{Conclusion}
\newtheorem{condition}[theorem]{Condition}
\newtheorem{conjecture}[theorem]{Conjecture}
\newtheorem{corollary}[theorem]{Corollary}
\newtheorem{criterion}[theorem]{Criterion}
\newtheorem{definition}[theorem]{Definition}
\newtheorem{example}[theorem]{Example}
\newtheorem{exercise}[theorem]{Exercise}
\newtheorem{lemma}[theorem]{Lemma}
\newtheorem{notation}[theorem]{Notation}
\newtheorem{problem}[theorem]{Problem}
\newtheorem{proposition}[theorem]{Proposition}
\newtheorem{remark}[theorem]{Remark}
\newtheorem{solution}[theorem]{Solution}
\newtheorem{summary}[theorem]{Summary}
\def\r{{\bf{r}}}
\def\i{{\bf{i}}}
\def\j{{\bf{j}}}
\def\m{{\bf{m}}}
\def\k{{\bf{k}}}
\def\kt{{\tilde{\k}}}
\def\mt{{\hat{t}}}
\def\mG{{\hat{G}}}
\def\mg{{\hat{g}}}
\def\mGa{{\hat{\Gamma}}}
\def\mS{{\hat{\Sigma}}}
\def\mT{{\hat{T}}}
\def\K{{\bf{K}}}
\def\P{{\bf{P}}}
\def\q{{\bf{q}}}
\def\Q{{\bf{Q}}}
\def\p{{\bf{p}}}
\def\x{{\bf{x}}}
\def\X{{\bf{X}}}
\def\Y{{\bf{Y}}}
\def\F{{\bf{F}}}
\def\G{{\bf{G}}}
\def\bG{{\bar{G}}}
\def\mbG{{\hat{\bar{G}}}}
\def\M{{\bf{M}}}
\def\V{\cal V}
\def\tchi{\tilde{\chi}}
\def\tx{\tilde{\bf{x}}}
\def\tk{\tilde{k}}
\def\tK{\tilde{\bf{K}}}
\def\tq{\tilde{\bf{q}}}
\def\tQ{\tilde{\bf{Q}}}
\def\si{\sigma}
\def\ep{\epsilon}
\def\hep{{\hat{\epsilon}}}
\def\al{\alpha}
\def\be{\beta}
\def\ep{\epsilon}
\def\bep{\bar{\epsilon}_\K}
\def\up{\uparrow}
\def\de{\delta}
\def\De{\Delta}
\def\up{\uparrow}
\def\dwn{\downarrow}
\def\ksi{\xi}
\def\etha{\eta}
\def\product{\prod}
\def\goto{\rightarrow}
\def\switch{\leftrightarrow}

\title{The Dynamical Cluster Approximation (DCA) versus the Cellular 
Dynamical Mean Field Theory (CDMFT) in strongly correlated electrons 
systems}
\author{
K.~Aryanpour,
Th.~A.~Maier and M.~Jarrell
}
\address{University of Cincinnati, Cincinnati OH 45221, USA}

\begin{abstract} We are commenting on the article Phys.\ Rev.\ {\bf B
	65}, 155112 (2002) by G.\ Biroli and G.\ Kotliar in which they
	make a comparison between two cluster techniques, the {\it
	Cellular Dynamical Mean Field Theory} (CDMFT) and the {\it
	Dynamical Cluster Approximation} (DCA). Based upon an incorrect
	implementation of the DCA technique in their work, they conclude
	that the CDMFT is a faster converging technique than the DCA. We
	present the correct DCA prescription for the particular model
	Hamiltonian studied in their article and conclude that the DCA,
	once implemented correctly, is a faster converging technique for
	the quantities averaged over the cluster. We also refer to their
	latest response to our comment where they argue that instead of
	averaging over the cluster, local observables should be
	calculated in the bulk of the cluster which indeed makes them
	converge much faster in the CDMFT than in the DCA. We however
	show that in their original work, the authors themselves use the
	cluster averaged quantities to draw their conclusions in favor
	of using the CDMFT over the DCA.    

\end{abstract}
\maketitle
In their article,\cite{biroli} G. Biroli and G. Kotliar compare two
cluster methods, the {\it Dynamical Cluster Approximation} (DCA) and
the {\it Cellular Dynamical Mean Field Theory} (CDMFT) for {\it the
simplified one-dimensional large-N model}\cite{affleck} and conclude
that the CDMFT converges faster with cluster size than the DCA.

This is a surprising result and in contradiction with exact results we
have previously published in Ref.~\cite{DCA_maier} where we have shown
that for large linear cluster sizes $L_c$, the DCA converges with
corrections of ${\cal O}(1/L_{c}^2)$ while the CDMFT converges with
corrections of ${\cal O}(1/L_{c})$. The objective of this comment is
to resolve this controversy.

The difference in scaling behaviors of the DCA and the CDMFT is a
direct consequence of their different boundary conditions of the
cluster. The DCA \cite{DCA_hettler1,DCA_hettler2} and the CDMFT
\cite{kotliar} are both approximative methods incorporating nonlocal
correlations in correlated electron systems relinquished by fully
local techniques such as the Dynamical Mean Field Approximation (DMFA)
\cite{metzner-vollhardt,muller-hartmann,review1,review2} and the
Coherent Potential Approximation (CPA). \cite{taylor} The DCA is
formulated in momentum space by dividing the first Brillouin zone into
a number of subcells. It reduces the complexity of the problem by {\it
coarse-graining} over all the momenta within a subcell. As a result,
the real lattice is mapped onto a cluster with periodic boundary
conditions coupled to a mean field. The CDMFT on the other hand is
formulated in real space and maps the lattice onto a cluster with open
boundary conditions.

We realized that the implementation of the DCA algorithm in
Ref.~\cite{biroli} is incorrect as they do not coarse-grain the
nonlocal interaction and their analysis lacks an accurate scaling
comparison in terms of the cluster size for these two methods. In
addition, instead of presenting physically relevant lattice quantities
they plot the corresponding cluster quantities which in the DCA in
general are not identical to their lattice counterparts. We repeat the
calculations of Ref.~\cite{biroli} by implementing the DCA correctly
and conclude that for any cluster size $L_c$, the DCA results are
closer to the exact solution of the studied toy model than the CDMFT
results, consistent with our previous findings. \cite{DCA_maier}

The exact mean field treatment of the {\it simplified one-dimensional
large-N model} Hamiltonian (Eq.~4 in Ref.~\cite{biroli}) yields
\begin{equation}
\label{eq:chi-exact}
E_{k}=-2(t+\chi)\cos k+\mu,\hspace{0.7cm}\chi=\frac{1}{L}\sum_{k}f(\beta 
E_{k})\cos k\,,
\end{equation}
as self-consistent equations for the dispersion $E_k$, with the number
of sites $L$, the chemical potential $\mu$, the inverse temperature
$\beta$ and the Fermi function $f(\beta E_{k})$. In the dispersion
$E_{k}$, one can derive the self-energy as
\begin{equation}
\label{eq:self-exact}
\Sigma(k)=-2~\chi~\cos k\,,
\end{equation}
where the $\cos k$ factor comes from the interaction between nearest
neighbors.

In the DCA, the first Brillouin zone is divided into $L_c$ subcells
with a linear size of $\Delta k=2\pi/L_c$. The irreducible quantities
are constructed from the coarse-grained propagators.  Both the Green
function, and the interaction (when it is non-local) must be coarse
grained\cite{DCA_hettler2,aryanpour}.  Thus, the self-energy part in
Eq.~\ref{eq:self-exact} changes to its DCA counterpart
\begin{equation}
\label{eq:self-DCA}
\Sigma_{DCA}(K_c)\approx-2\chi^{DCA}_{cl}\overline{\cos (K_c)}\,,
\end{equation}
where
\begin{equation}
\label{eq:CG-cos}
\overline{\cos (K_c)}=\frac{L_c}{L}\sum_{\tk}\cos (K_c+\tk) \,,
\end{equation}
comes from coarse-graining the nearest neighbor interaction over all
the momenta $\tk$ within each subcell around the cluster momenta $K_c$
representing the subcells. G.~Biroli and G.~Kotliar in
Ref.~\cite{biroli} (Eq.~6 for the dispersion) do not coarse-grain
$\cos k$ in the nonlocal interaction and instead, they just substitute
it by $\cos K_c$. This implementation of the DCA is incorrect and
overestimates the effects of the interaction. The correct
self-consistent DCA equations for the dispersion are
\begin{eqnarray}
\label{eq:chi-DCA}
\hspace{0.0cm}\chi^{DCA}_{cl}=\frac{1}{L_c}\sum_{\K_c}\cos 
(K_c)\times\overline{f(\beta 
E(K_c))},\nonumber\\&&\hspace*{-7.0cm}\overline{f(\beta 
E(K_c))}=\frac{L_c}{L}\sum_{\tk}f(\beta 
E_{K_c+\tk}),\nonumber\\&&\hspace*{-7.0cm}E_{K_c+\tk}\approx-2~t~\cos 
(K_c+\tk)-2\chi^{DCA}_{cl}\overline{\cos (K_c)}+\mu.
\end{eqnarray}
After self-consistency is achieved, one can use the cluster quantity 
$\chi^{DCA}_{cl}$ to compute its lattice counterpart
\begin{equation}
\label{eq:chi-latt}
\chi^{DCA}_{latt}=\frac{1}{L}\sum_{k}f(\beta E_{K_c+\tk})\cos k \,.
\end{equation}

G.~Biroli and G.~Kotliar plot the cluster quantity $\chi_{cl}^{DCA}$
  instead of the physically relevant lattice quantity
  $\chi^{DCA}_{latt}$ (Fig.~1 in Ref.~\cite{biroli}). In the DCA,
  non-local cluster quantities like $\chi_{cl}^{DCA}$ are not identical
  to their corresponding lattice quantities and thus have no physical
 meaning. In the CDMFT, since there exists one to one correspondence
  between the lattice quantity $\chi^{CDMFT}_{latt}$ and the cluster
  quantity $\chi^{CDMFT}_{cl}$, they are  identical.

By numerically solving the self-consistent Eqs.~\ref{eq:chi-exact} and
\ref{eq:chi-DCA} and the one for the CDMFT (Eq.~8 in
Ref.~\cite{biroli}), we can make a comparison between the exact
solution (cf. Eq.~\ref{eq:chi-exact} in this article) and the
predictions by the DCA and the CDMFT.
\begin{figure}
\includegraphics*[width=3.4in]{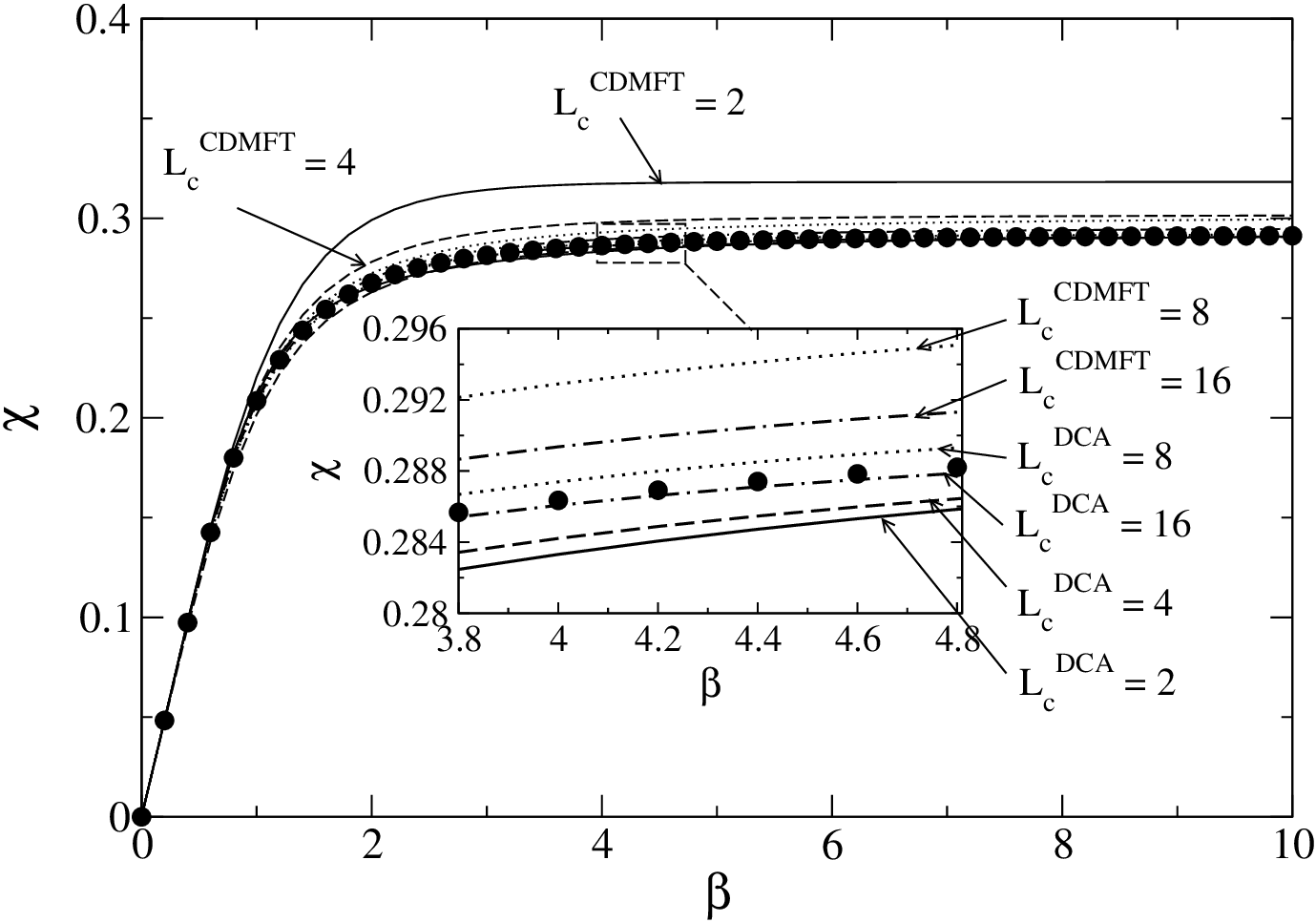}
\vspace{0.01cm}
\caption[a]{$\chi$ as a function of $\beta$ (inverse temperature) at 
$t=1$ and filling $n=0.37$ ($\mu\approx1$) for both the DCA and the 
CDMFT at $L_c=2, 4, 8, 16$. The exact result is represented by the solid 
circles.}
\label{xi.vs.beta}
\end{figure}

\begin{figure}
\hspace{-0.4cm}
\includegraphics*[width=3.5in]{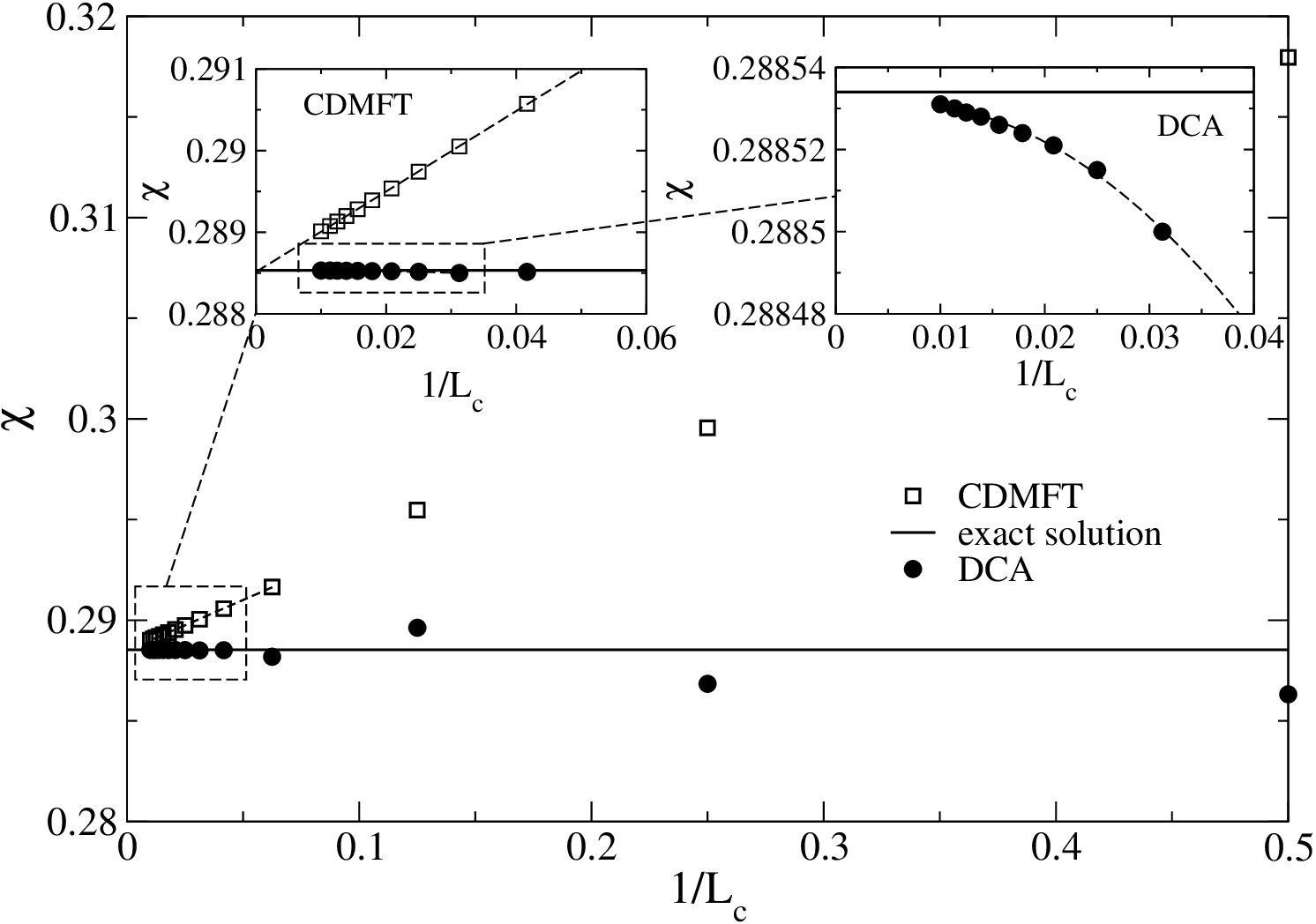}
\vspace{0.03cm}
\caption[a]{$\chi$ as a function of $1/L_c$ at $\beta=5$, $t=1$ and
filling $n=0.37$ ($\mu\approx1$). The insets show the convergence of
the DCA (filled circles) and CDMFT (open squares) results to the exact
solution (solid line) beyond $L_c=16$ for the CDMFT and $L_c=32$ for
the DCA. The dashed lines in the insets represent the linear
and quadratic fits in $1/L_c$ to the CDMFT and DCA results
respectively.}
\label{xi.vs.size}
\end{figure}
\begin{figure}
\hspace{-0.4cm}
\includegraphics*[width=3.4in]{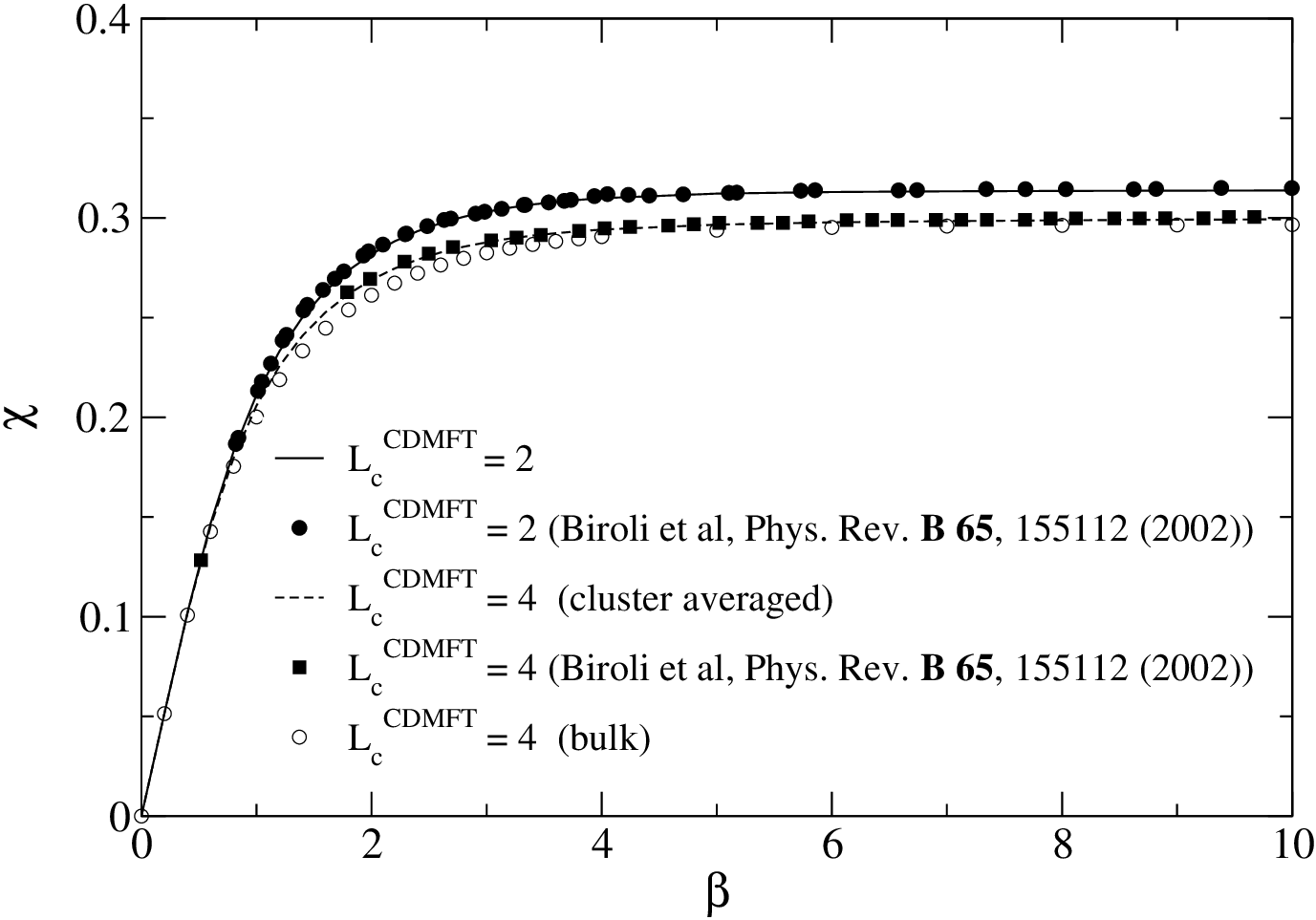}
\vspace{0.03cm}
\caption[a]{$\chi$ as a function of $\beta$ (inverse temperature) at 
$t=1$ and $\mu=1$ for the CDMFT at $L_c=2, 4$. The second (filled
circles) and forth (filled squares) curves are the digitized results
extracted from Fig.~1 in Ref.~\cite{biroli} plotted versus the CDMFT
results obtained in this work (first and third curves). The last curve
(open circles) refers to $\chi$ at $L_c=4$ in the bulk (center of the
cluster).}
\label{comp.of.xi}
\end{figure}
In Fig.~\ref{xi.vs.beta}, we plot $\chi$ versus $\beta$ for $t=1$ and
fixed filling of $n=0.37$ ($\mu\approx1$) using the correct
implementation of the DCA (compared to the results of Fig.~1 in
Ref.~\cite{biroli}) and CDMFT for $L_c=2, 4, 8$ and $16$. It is more
physical to do the calculations at fixed filling $n$ instead of fixed
chemical potential $\mu$ as it was done in Ref.~\cite{biroli}. As
illustrated, it is clear that the DCA results converge faster than the
CDMFT ones to the exact solution even for small $L_c$. The CDMFT
results converge consistently whereas for small $L_c$ ($L_c=8$ in
particular) the DCA results fluctuate around the exact
solution. Nevertheless, at large enough $L_c$, the DCA behaves
consistently and converges from below. We have avoided odd cluster
sizes in these calculations because for the DCA, the periodic boundary
conditions on the cluster suppress $Q=\pi$ fluctuations by frustrating
the cluster.

In Fig.~\ref{xi.vs.size}, we plot $\chi$ as a function of the inverse 
cluster size $1/L_c$ computed by the DCA and the
CDMFT at fixed inverse temperature $\beta=5$ for $t=1$ and $n=0.37$
($\mu\approx1$) and compare the results with the exact solution (solid
line). The DCA begins to converge quadratically and the CDMFT linearly
in $1/L_c$ to the exact solution beyond $L_c=32$ and $L_c=10$
respectively (the fits in the insets). The scaling behavior of both
the DCA and CDMFT at large enough $L_c$ are consistent with our
previous results. \cite{DCA_maier}

It is noteworthy that for this Hamiltonian in particular, the values
of $L_c$ beyond which the scaling behaviors of the DCA and the CDMFT
are observed are relatively high. E.g., in the one dimensional
Falicov-Kimball model at half filling, these scaling behaviors are
manifest beyond $L_c=2$. \cite {DCA_maier}

The results in both Figs.~\ref{xi.vs.beta} and \ref{xi.vs.size}
indicate the superiority of the DCA over the CDMFT in terms of faster
convergence with the cluster size $L_c$. The
DCA also significantly reduces the computational CPU time compared to
the CDMFT. In the CDMFT, because of the open boundary conditions on
the cluster, the Green function in both real and momentum spaces is
not diagonal. Thus, expensive diagonalizations of large matrices are
required for large cluster sizes.

In their latest response to this comment, \cite{biroli.resp} G.\ Biroli
and G.\ Kotliar argue that in the CDMFT, whenever the correlation length
is finite, {\it local observables} such as $\chi$ in this problem are
best estimated  in the bulk than near the boundaries of the cluster. In
fact, their values in the bulk converge exponentially in the CDMFT while
in the DCA they still converge as $1/L_{c}^2$. Because the CDMFT breaks
the translational invariance inside the cluster, local observables
calculated in the bulk of the cluster differ from those on the boundary.
Provided that the system is far from a transition, the sites in the
center of the CDMFT cluster couple to the mean-field only through
propagators which fall exponentially with distance. Local observables,
when extracted from the center of the cluster instead of from their flat
average, hence converge much faster, i.e. exponentially as opposed to
$1/L_c$. As a matter of fact, this point was made previously by some of us 
\cite{maier:dca4}; however, for most models, it is difficult to formulate 
a causal self energy from quantities measured on the center of the cluster. 
Nevertheless, in their original work in Fig.1 of Ref.~\cite{biroli}, 
G.\ Biroli and G.\ Kotliar themselves perform a flat averaging of the 
local observable $\chi$ and by comparing its results
with their incorrect DCA results, they conclude that {\it the CDMFT
converges better}. To prove that the results in Fig.1 in
Ref.~\cite{biroli} are indeed the flat averaged $\chi$ and not its value
in the bulk, we plot in Fig.~\ref{comp.of.xi} both the flat averaged and
bulk $\chi$ versus $\beta$ for $L_c=2, 4$. We have also extracted the
CDMFT curves in Fig.1 of Ref.~\cite{biroli} for $L_c=2, 4$ (filled
circles and squares respectively) using {\it Engauge} digitizing package. Their
perfect match with the flat averaged curves and their deviation from the
bulk result for $L_c=4$ best indicate that
$\chi$ in Fig.1 of Ref.~\cite{biroli} was flat averaged and not
calculated in the bulk.

Another significant difference between these two techniques appears in
the calculation of the self-energy. The DCA approximates the lattice
self-energy by a constant within a DCA subcell in momentum space and
as a result, the self-energy becomes a step function in
$k$-space. Thus, one may use an approach such as the Akima spline
which is the smoothest curve interpolating through these step
functions to generate an analytic form for the self-energy. Using the
Akima spline is consistent with the assumption that the self-energy is
a smoothly varying function in momentum space. In the CDMFT, since the
lattice self-energy is approximated by the cluster self-energy in real
space, its Fourier transform to momentum space yields a smooth
self-energy curve. Due to the open boundary conditions of the cluster,
the translational invariance is violated and the self-energy in
momentum space is not diagonal. Nevertheless, by using the {\it Random
Phase Approximation} (RPA), \cite{senechal} the translational
invariance can be restored and the self-energy becomes a diagonal
function in momentum space.  

\begin{acknowledgments}
We would like to acknowledge M.~H.~Hettler and H.~R.~Krishnamurthy for 
the review of this comment and useful suggestions. This work was supported 
by NSF grants DMR-0073308 and DMR-9704021.
\end{acknowledgments}

\end{document}